\documentclass[12pt,english]{article}
\usepackage{lmodern}

\usepackage[T1]{fontenc}
\usepackage[utf8]{inputenc}
\usepackage{geometry}
\usepackage{amsmath}
\usepackage{graphicx}
\usepackage{setspace}
\makeatletter

\providecommand{\tabularnewline}{\\}

\newcommand{\lyxaddress}[1]{
	\par {\raggedright #1
	\vspace{1.4em}
	\noindent\par}
}
\def\be{\begin{equation}}
\def\ee{\end{equation}}

\def\bea{\begin{eqnarray}}
\def\eea{\end{eqnarray}}

\def\ba{\begin{array}}
\def\ea{\end{array}}

\def\bc{\begin{center}}
\def\ec{\end{center}}

\def\bl{\begin{flushleft}}
\def\el{\end{flushleft}}

\def\br{\begin{flushright}}
\def\er{\end{flushright}}

\def\bi{\begin{itemize}}
\def\ei{\end{itemize}}

\def\bt{\begin{tabular}}
\def\et{\end{tabular}}

\numberwithin{equation}{section}

\usepackage[winfonts,UTF8]{ctex}
\usepackage{slashed}
\usepackage{hyperref}
\usepackage{graphicx}
\usepackage[numbers,sort&compress]{natbib}
\date{}

\hypersetup{
colorlinks=true,
linkcolor=black
}

\makeatother

\usepackage{babel}
\begin{document}
\title{Greybody factor and power spectra of the Hawking radiation in the novel $4D$ Einstein-Gauss-Bonnet de-Sitter gravity}
\author{Cheng-Yong Zhang$^{1}$, Peng-Cheng Li$^{2,3}$, Minyong Guo$^{2*}$}
\maketitle

\lyxaddress{\begin{center}
\textit{1.Department of Physics and Siyuan Laboratory, Jinan University,
Guangzhou 510632, China}\\
\textit{2. Center for High Energy Physics, Peking University, No.5
Yiheyuan Rd, Beijing 100871, P. R. China}\\
\textit{3. Department of Physics and State Key Laboratory of Nuclear
Physics and Technology, Peking University, No.5 Yiheyuan Rd, Beijing
100871, P.R. China}
\par\end{center}}

\begin{abstract}
We investigate Hawking radiation in the novel $4D$ Einstein-Gauss-Bonnet gravity which is recently  formulated by Glavan and Lin [Phys. Rev. Lett. 124, 081301 (2020)]. We first find the constraints on $\alpha$ and $\Lambda$ to retain a dS black hole in the spacetime. Both the greybody factor and the power spectra of the Hawking radiation of the massless scalar are studied for the full range of various parameters including the GB coupling constant $\alpha$, the cosmological constant $\Lambda$ and the constant related to the scalar $\xi$ numerically. In particular, we find the greybody factor is larger for a negative $\alpha$ than the case for $\alpha\ge0$. While, the power spectra of the Hawking radiation is higher for a allowed negative $\alpha$ than others with non-negative $\alpha$. The reason is that the temperature of the black hole would be very high when $\alpha<0$, even when $\alpha$ approaches the lower bound, the temperature would be arbitrary high.
\end{abstract}

\vfill{\footnotesize Email: zhangcy@email.jnu.edu.cn,\,\, lipch2019@pku.edu.cn,\,\, minyongguo@pku.edu.cn. \\  \hspace*{1.5em}  $^*$Corresponding author}

\maketitle
\newpage
\section{Introduction}

Hawking radiation is one of the most important discoveries in quantum gravity, although it was derived semi-classically \cite{Hawking:1974sw} in virtue of quantum field theory in cured spacetime. Since then, people believed black holes do have temperatures and satisfy the laws of thermodynamics \cite{Hawking:1976de}. On the other hand, Hawking radiation plays a central role in the study of the black hole information paradox, which may be the toughest obstacle in the way to understanding quantum gravity thoroughly \cite{Hawking:1976ra}. Based on the awareness of its importance and necessity, a lot of significant progress has been made in the previous studies both experimentally and theoretically. On the experimental side, it was found that the Hawking radiation of tiny black holes may be observed by particle colliders through LHC \cite{Argyres:1998qn,Giddings:2001bu,Dimopoulos:2001hw,Kanti:2008eq}, while Hawking radiation is too small to be directly detected through astronomical observations. Instead, Hawking radiation was proved to work in the laboratory analogues in terms of the theory of analogue gravity \cite{Unruh:1980cg} . Further, some analogues of black-hole horizons in the laboratory have been realized and many evidences came into being supporting the universality of Hawking radiation \cite{Steinhauer:2014dra,Steinhauer:2015saa}, for more complete reviews see Ref. \cite{Barcelo:2005fc}. On the purely theoretical side, Hawking radiation also has drawn attentions on many aspects of gravity mainly including quantum information theory of black hole \cite{Harlow:2014yka}, holographic dual in the frame of AdS/CFT correspondence \cite{Maldacena:1997re} and Hawking emission spectrum described in terms of the greybody factor in general relativity, extra-dimension models and other alternative gravity theory \cite{ Page1976, Das1997, Brady1997, Harris2003, %Kanti2004, 
Kanti2005, Grain2005, Konoplya2010, Harmark2010, Crispino2013, Zhu2014, Kanti2014, %%bJorge2015, bDong2015,bSporea2016, bPanotopoulos2017_2, Ovgun:2018gwt,
Brito2015,Zhang2014a,Zhang2014b, Pappas2016, bAhmed2016, bPanotopoulos2017,bMiao2017, Kanti2017, Pappas2017, bKuang2017, Zhang2017,  Yekta:2018hye, Xu:2018liy, Li2019, Volkel:2019ahb, Konoplya:2019hml, Xu:2019krv, Konoplya:2019hlu, Konoplya:2019ppy}.

Thereinto, the Einstein-Gauss-Bonnet (EGB) gravity is known to be one of the most promising candidates for modified gravity \cite{Clifton2011}. A well-defined solution for any dimension ($D\ge5$) has been found \cite{Boulware1985} and many non-trivial effects from the Gauss-Bonnet term on the black hole solution appear indicating the study of EGB gravity not only helps us understand black holes and black branes more deeply \cite{Cai:2001dz,Konoplya2008, Cuyubamba2016, Konoplya2017} but also advances the development of the holographic gravity in the AdS/CFT correspondence \cite{Buchel:2009sk}. Among these studies,
many important results related to Hawking radiation in EGB gravity were also obtained  and aroused  a lot of concern and discussion \cite{%Kanti2005,
Grain2005, Konoplya2010}. Despite the achievements,  there are some localization and regrettable in the study: the first of these may be the
deficiencies that none of these results can be compared directly with real black holes since the solutions in EGB gravity is limited to greater than or equal to five dimensions.

Recently, a novel 4D EGB gravity was formulated by rescaling the GB coupling constant which completed the missing piece of the EGB gravity. What is even more exciting, a spherically symmetric black hole solution in this theory was derived by D. Glavan and C. Lin \cite{ Glavan2019} is free of the singularity problem, which is supposed to be done by considering quantum corrections \cite{Cai:2009ua, Tomozawa:2011gp, Cognola:2013fva}. Therefore, this novel 4D EGB gravity has aroused great interest since its publication \cite{Konoplya:2020bxa,Guo:2020zmf,Fernandes:2020rpa,Casalino:2020kbt,Konoplya:2020qqh,Wei:2020ght,Kumar:2020owy,
Hegde:2020xlv,Ghosh:2020vpc,Doneva:2020ped,Zhang:2020qew,Lu:2020iav,Singh:2020xju}. It's worth mentioning that, in \cite{Fernandes:2020rpa}, the authors first discussed the negative GB coupling constant in details and found a negative GB coupling constant is allowed to retain a black hole, and some appealing features from the GB term with a negative coupling constant were obtained. However, one doesn't know yet if a negative coupling constant works when a positive cosmological constant is added in the novel 4D EGB gravity, furthermore, the works on Hawking radiation are also absent.

In this paper, we focus on the 4D spherically symmetric EGB-dS black hole solution and investigate the full range of the GB coupling constant to retain a dS black hole. Because of the positive cosmological constant, we found a negative GB coupling constant is also allowed but both the GB coupling constant and cosmological constant are constrained non-trivially. After having the preparation, we have a completed study on the greybody factor of the Hawking radiation firstly and move to  the power spectra of the Hawking radiation of the massless scalar in the 4D GB-dS background. The effects of the GB coupling constant and cosmological constant on both aspects of Hawking radiation are discussed as well as the coupling constant related to the scalar field. In particular, some new features are found compared with the previous results form Einstein-Gauss-Bonnet (EGB) gravity when the GB coupling constant is negative.

The paper is organized as follows. In section \ref{sec:Background}, we give a short review on the novel 4D EGB gravity and find the full range of the GB coupling constant and cosmological constant when the spacetime contains a dS black hole. In section \ref{seccfp}, we introduce the scalar field perturbation. Next, we turn our attention to the greybody factor in section \ref{sec:Greybody-factor}. And the energy emission rate of Hawking radiation are discussed in section 5. We close our paper with a conclusion in section \ref{sec:Summary}.

\section{The novel 4-dimensional EGB-dS black hole solution\label{sec:Background}}

The Einstein-Gauss-Bonnet gravity with a positive cosmological constant
in $D$-dimensional spacetime has the action of the form
\begin{equation}
S_{G}=\frac{1}{16\pi G}\int d^{D}x\sqrt{-g}\left[R-2\Lambda+\alpha\mathcal{L}_{GB}\right],\label{eq:EGBAction}
\end{equation}
where $G$ is the $D$-dimensional Newton's constant, $R$ is the
Ricci scalar, $\Lambda$ is the cosmological constant, and $\alpha$
is the Gauss-Bonnet (GB) coupling constant of dimension $(length)^{2}$.
The Gauss-Bonnet term
\begin{equation}
\mathcal{L}_{GB}=R_{\mu\nu\rho\sigma}R^{\mu\nu\rho\sigma}-4R_{\mu\nu}R^{\mu\nu}+R^{2}.
\end{equation}
 It is a total derivative in four dimensional spacetime and thus has
no contribution to the dynamics in general. Surprisingly, novel black
hole solutions were discovered recently in this theory by rescaling
the coupling constant as
\begin{equation}
\alpha\to\frac{\alpha}{D-4}.
\end{equation}
In the limit $D\to4$, the Lovelock's theorem is circumvented and
new black hole solution was found in \cite{Glavan2019}.

The novel spherically symmetric GB-dS black hole solution of (\ref{eq:EGBAction})
in four dimensional spacetime can be described by
\begin{align}
ds^{2} & =-fdt^{2}+\frac{dr^{2}}{f}+r^{2}d\Omega_{2}^{2},\label{eq:4metric}\\
f & =1+\frac{r^{2}}{2\alpha}\left(1-\sqrt{1+\frac{4\alpha m}{r^{3}}+\frac{4\alpha\Lambda}{3}}\right).\nonumber
\end{align}
 Here $d\Omega_{2}^{2}$ is the line element of the $2$-dimensional
unit sphere $S^{2}$. The parameter $m$ is related to the black hole
mass $M$ by $m=2GM$. This solution has the same form as their higher
dimensional companions. But there are important differences. The four
dimensional solution (\ref{eq:4metric}) can have three horizons in
some parameter region, while the higher dimensional solutions have
at most two horizons. The center singularity of (\ref{eq:4metric})
is repulsive, while it is attractive in higher dimensions. Thus the
four dimensional solution is free from the singularity problem. Motivated
by these differences, we study the Hawking radiation of novel solution
in this paper.

\begin{figure}[h]
\begin{centering}
{\footnotesize{}\includegraphics[scale=0.8]{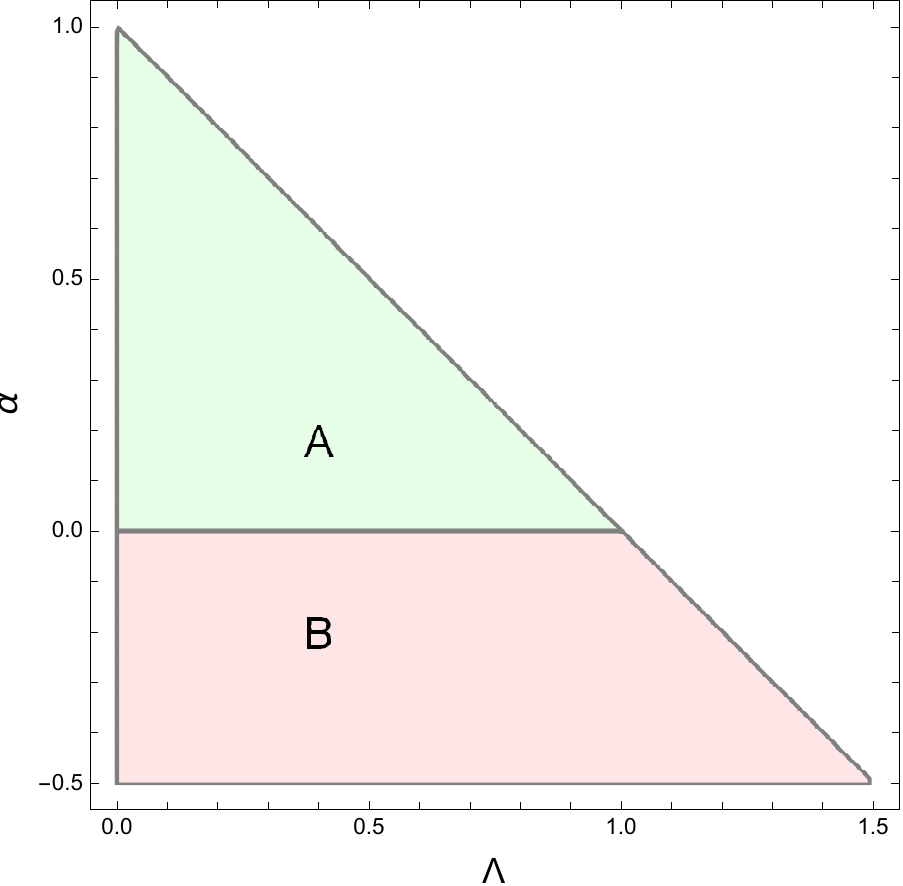}}{\footnotesize\par}
\par\end{centering}
{\footnotesize{}\caption{\label{fig:ParameterRegion} The parameter region where allows the
event horizon and cosmological horizon. The region is bounded by $\alpha>-0.5$,
$\Lambda>0$ and $\alpha+\Lambda<1$. The metric function is real for
all $r>0$ in region $A$, while becomes complex for small $r$ in
region $B$.}
}{\footnotesize\par}
\end{figure}

We fix $r_{h}=1$ for convenience. Then the mass parameter $m$ can
be expressed as
\begin{equation}
m=1+\alpha-\frac{\Lambda}{3}>0.
\end{equation}
 The free parameters of the background are $\alpha,\Lambda$ now.
In this paper, we focus on the parameter region where both the black hole event horizon and cosmological horizon exist. In addition we find
\be
f(\infty)=-\infty,\quad\quad f(0^+)=1,\quad\quad f^\prime(0^+)<0
\ee
for $\alpha>0$ while for $\alpha<0$, one has
\bea
f(\infty)=-\infty,\quad\quad 0<r_0<1
\eea
where $r_o$ is the root when the quadratic radical of the metric function $f(r)$ is vanishing and we have used the condition $m>0$ for the inequalities. By using these conditions in combination, we conclude $f(1)=0$ and $f^\prime(1)>0$ give the necessary and sufficient condition that this spacetime always contains a dS black hole, that is, both the event horizon and the cosmological horizon exist. Thus, we find the allowed region and show it in Fig. \ref{fig:ParameterRegion}.

{\footnotesize{}}

\section{The scalar field perturbation}\label{seccfp}

We focus on the Hawking radiation of a massless scalar field $\Phi$
in the background (\ref{eq:4metric}). Field $\Phi$ couples to gravity
minimally or non-minimally with coupling constant $\xi$. The corresponding
equation of motion is \cite{Zhang2017,Li2019}
\begin{equation}
\nabla_{\mu}\nabla^{\mu}\Phi=\xi R\Phi.\label{eq:KG}
\end{equation}
 For fluctuations of order $\mathcal{O}(\epsilon)$, the induced changes
of the spacetime geometry is at order $\mathcal{O}(\epsilon^{2})$.
Thus the effect of $\Phi$ on the background spacetime is negligible
\cite{Brito2015}. The above equation can be solved in a fixed background
given by (\ref{eq:4metric}).

For stationary spherically symmetric background, the scalar wave function
can be decomposed as
\begin{equation}
\Phi(t,r,\Omega)=\int d\omega\sum_{lm}e^{-i\omega t}\phi(r)Y_{lm}(\Omega),
\end{equation}
 where $Y_{lm}(\Omega)$ are spherical harmonics of the scalar wave
function on $S^{2}$. The angular and radial part are decoupled and
the radial master equation reads
\begin{equation}
\frac{1}{r^{2}}\frac{d}{dr}\left(fr^{2}\frac{d\phi}{dr}\right)+\left[\frac{\omega^{2}}{f}-\frac{l(l+1)}{r^{2}}-\xi R\right]\phi=0,\label{eq:RadialEq}
\end{equation}
 in which the Ricci scalar can be written as
\begin{equation}
R=-\partial_{r}^{2}f+\frac{2}{r^{2}}\left(-2r\partial_{r}f+1-f\right).
\end{equation}
The equation (\ref{eq:RadialEq}) can be written in the Schrödinger-like
form
\begin{equation}
\frac{d^{2}u}{dr_{\ast}^{2}}+(\omega^{2}-V_{\text{eff}})u=0,
\end{equation}
 where $r_{\ast}$ is the tortoise coordinate defined by $dr_{\ast}=\frac{1}{f}dr$,
and the new variable $u(r)=r\phi(r)$ is introduced. The effective
potential
\begin{equation}
V_{\text{eff}}=f\left(\frac{l(l+1)}{r^{2}}+\xi R+\frac{1}{r}\partial_{r}f\right).\label{eq:effPotential}
\end{equation}
It vanishes at the horizons of the spacetime and has a potential barrier
between the event horizon $r_{h}$ and the cosmological horizon $r_{c}$.
It encodes the information of the background and the field $\Phi$,
such as the angular momentum number $l$, scalar coupling constant
$\xi$, cosmological constant $\Lambda$ and GB coupling constant
$\alpha$. We will study the effects of these parameters on the Hawking
radiation in detail in this paper.

The radial equation has asymptotic behavior
\begin{align}
u\to & \begin{cases}
\mathcal{T}e^{-i\omega r_{\ast}}+\mathcal{O}e^{i\omega r_{\ast}}=\mathcal{T}(r-r_{h})^{i\frac{\omega}{\kappa_{h}}}+\mathcal{O}(r-r_{h})^{-i\frac{\omega}{\kappa_{h}}}, & r\to r_{h},\\
\mathcal{R}e^{-i\omega r_{\ast}}+\mathcal{I}e^{i\omega r_{\ast}}=\mathcal{R}(r-r_{c})^{i\frac{\omega}{\kappa_{c}}}+\mathcal{I}(r-r_{c})^{-i\frac{\omega}{\kappa_{c}}}, & r\to r_{c},
\end{cases}\label{eq:asymptoticBehavior}
\end{align}
near the event horizon and cosmological horizon. Factors $\kappa_{h}$
and $\ensuremath{\kappa_{c}}$ are the surface gravity on $r_{h}$
and $r_{c}$, respectively. The coefficients $\mathcal{I},\mathcal{R}$
and $\mathcal{T}$ are the amplitudes of the incident, reflected and
transmitted waves, respectively. The $\mathcal{O}$ term describes
an outgoing wave at the event horizon and will be set to zero hereafter.
Once these coefficients are worked out, we can get the greybody factor
of the Hawking radiation through the definition
\begin{equation}
|\gamma_{\omega l}|^{2}=1-|\mathcal{R}|^{2}/|\mathcal{I}|^{2}.\label{eq:GreyFactor}
\end{equation}

In general, the radial equation is hard to solve analytically. We
thus turn to solve the radial equation numerically. We impose the
ingoing boundary condition near the event horizon $r_{h}$ and integrate
the radial equation (\ref{eq:RadialEq}) towards the cosmological
horizon $r_{c}$. By comparing with the asymptotic behavior (\ref{eq:asymptoticBehavior})
near $r_{c}$, we can get the coefficients $\mathcal{R}$, $\mathcal{I}$
and therefore the greybody factor (\ref{eq:GreyFactor}). Various
numerical methods exist to solve the radial equation. Here we adopt
the method developed in \cite{Kanti2014,Pappas2016,Zhang2017,Li2019}.

\section{The greybody factor of Hawking radiation\label{sec:Greybody-factor}}

The free parameters are $\alpha,\Lambda$ and $\xi,l$. We study their
effects on the greybody factor in detail in this section.

\subsection{\label{subsec:AlphaL}Effects of $\alpha$ and $l$ on the greybody
factor }

When $\alpha\to0$, the novel solution (\ref{eq:4metric}) reduces
to the Schwarzschild-dS (SdS) black hole. As $\alpha$ increases,
the geometry is changed significantly due to the appearance of an
inner horizon in the black hole. One may expect that the greybody
factor of the Hawking radiation will also be affected significantly.

\begin{figure}[h]
\begin{centering}
{\footnotesize{}}%
\begin{tabular}{cc}
{\footnotesize{}\includegraphics[scale=0.8]{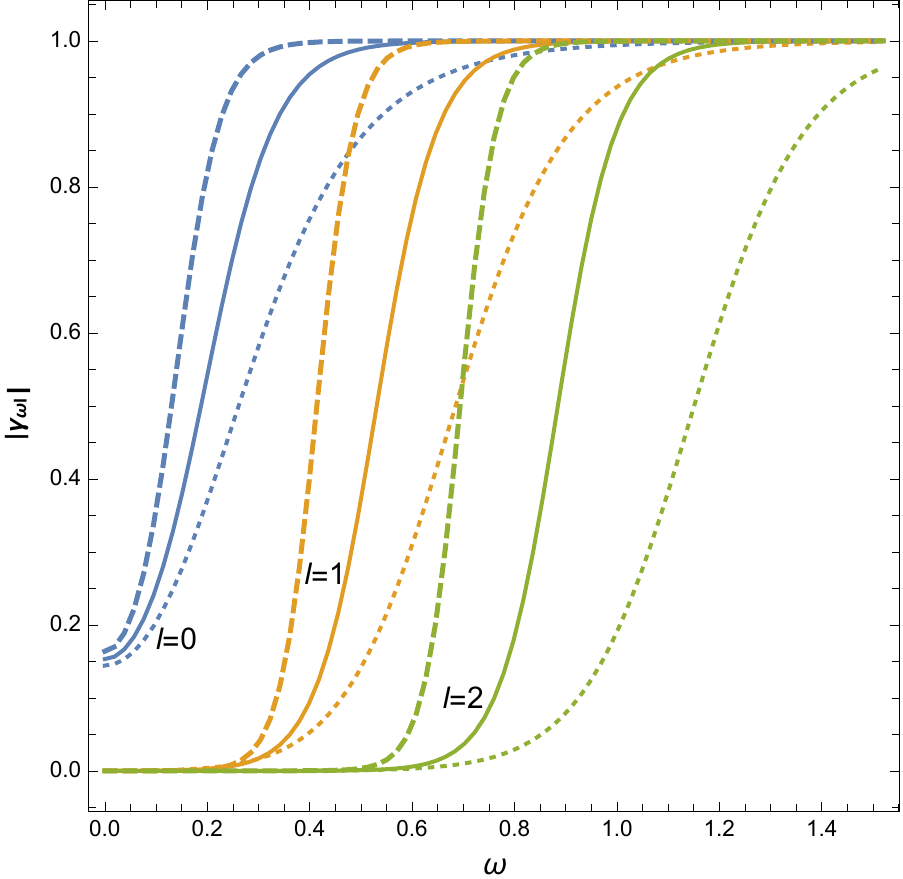}} & {\footnotesize{}\includegraphics[scale=0.8]{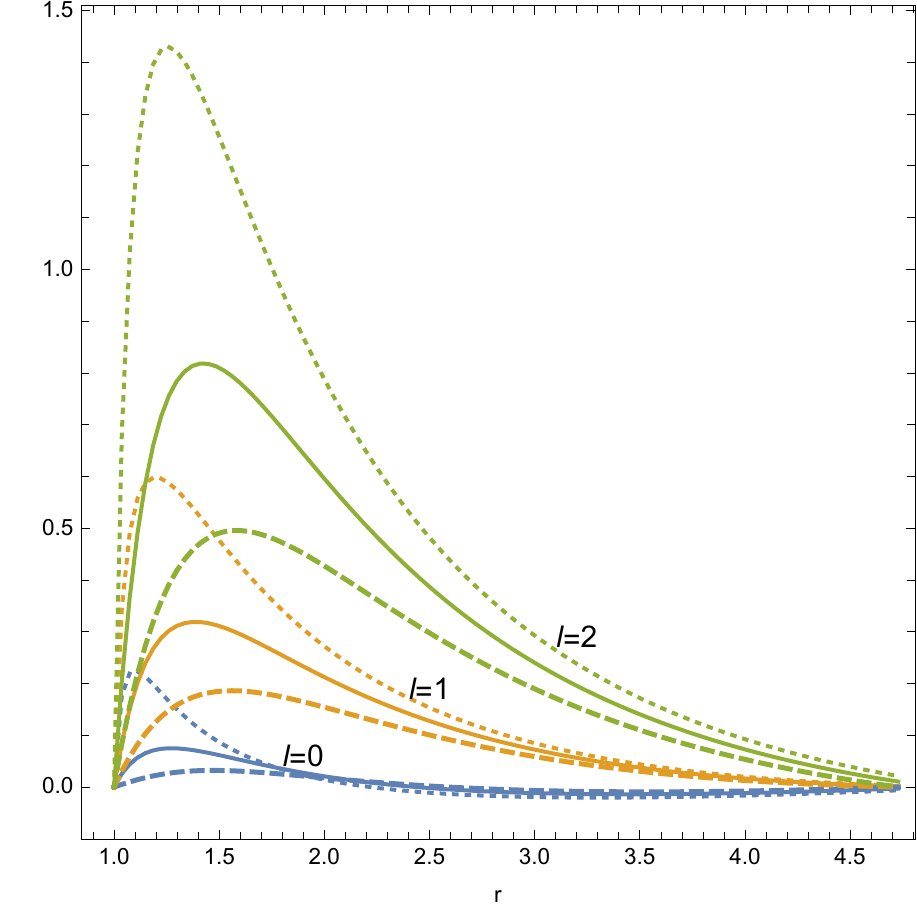}}\tabularnewline
\end{tabular}{\footnotesize\par}
\par\end{centering}
{\footnotesize{}\caption{\label{fig:AlphaL} Effects of $\alpha$ on the greybody factor
for different modes $l$ (left) and the corresponding effective potential
(right). Dotted lines for $\alpha=-0.3$. Solid lines for $\alpha=0$
(SdS), dashed lines for $\alpha=0.3$. We fix $\Lambda=0.1$, $\xi=0$
here.}
}{\footnotesize\par}
\end{figure}

We show the effects of $\alpha$ on the greybody factor for
different modes $l$ in the left panel of Fig. \ref{fig:AlphaL}.
The greybody factor is enhanced for all modes $l$ as $\alpha$ increases,
and is suppressed by $l$ when other parameters are fixed. This phenomenon
can be understood intuitively from the viewpoint the effective potential,
which is shown in the right panel of Fig. \ref{fig:AlphaL}. The larger
$l$, the higher potential barrier. The wave is more likely to be
reflected and leads to a smaller greybody factor according to (\ref{eq:GreyFactor}).
On the other hand, the potential barrier decreases with $\alpha$
when other parameters are fixed. The wave is unlikely to be reflected
and lead to a larger greybody factor. In fact, it can be shown numerically
that the maximum of the effective potential increases monotonically
with $l$ and decreases monotonically with $\alpha$. Thus $l$ always
suppress the greybody factor and $\alpha$ always enhance it. We would like to stress that compared to the Schwarzschild-dS black hole ($\alpha=0$), a positive $\alpha$ make the peak of the effective potential smaller, while a negative $\alpha$ increases the peak of the effective potential, which implies the exist of a negative $\alpha$ would make it harder for Hawking radiation to get through the barrier. This is a new phenomenon for us that one may guess black holes may be harder to evaporate than we thought if we focus on the greybody factor. But this is not true since one has to turn to the energy emission rate which will be investigated in the next section.

{\footnotesize{}}

Note that for the dominant mode $l=0$, the greybody factor does not
vanish when $\omega\to0$. This is a characteristic feature of the
minimally coupled massless scalar propagating in the dS spacetime.
The presence of an $\alpha$ does not change this feature qualitatively.
In the following subsection, we will see that this feature is changed
qualitatively for nonminimally coupled scalar.

\subsection{\label{subsec:XiL}Effect of $\xi$ on the greybody factor }

The effects of $\xi$ on the greybody factor is shown in Fig. \ref{subsec:XiL}.
We see that $\xi$ decreases the greybody factor when other parameters
are fixed. Similarly, this phenomenon can also be understood intuitively
from the effective potential. Now we know that positive $\alpha$
increases the greybody factor, while $\xi$ decreases it. There must
be competition between $\alpha$ and $\xi$. However, we find that
the effect of $\alpha$ is much weaker than the effect of $\xi$.
In the presence of $\alpha$, the derivative of the maximum of the
effective potential with respect to $\xi$ depends on the sign of
curvature $R$, as can be seen from (\ref{eq:effPotential}). To increase
the greybody factor, this derivative must be negative. This is possible
only when $\Lambda<3/56$. Above this threshold, the greybody factor
always decreases with $\xi$ in the whole allowed range of $\alpha$.
This phenomenon exists also in higher dimensional spacetimes. For
example, the threshold is $\Lambda<1/7$ when $D=5$ and $\Lambda<40/143$
when $D=6$. The threshold is larger in higher dimensions, which implies
that the effect of $\alpha$ gets stronger in higher dimension.

{\footnotesize{}}
\begin{figure}[h]
\begin{centering}
{\footnotesize{}\includegraphics[scale=0.8]{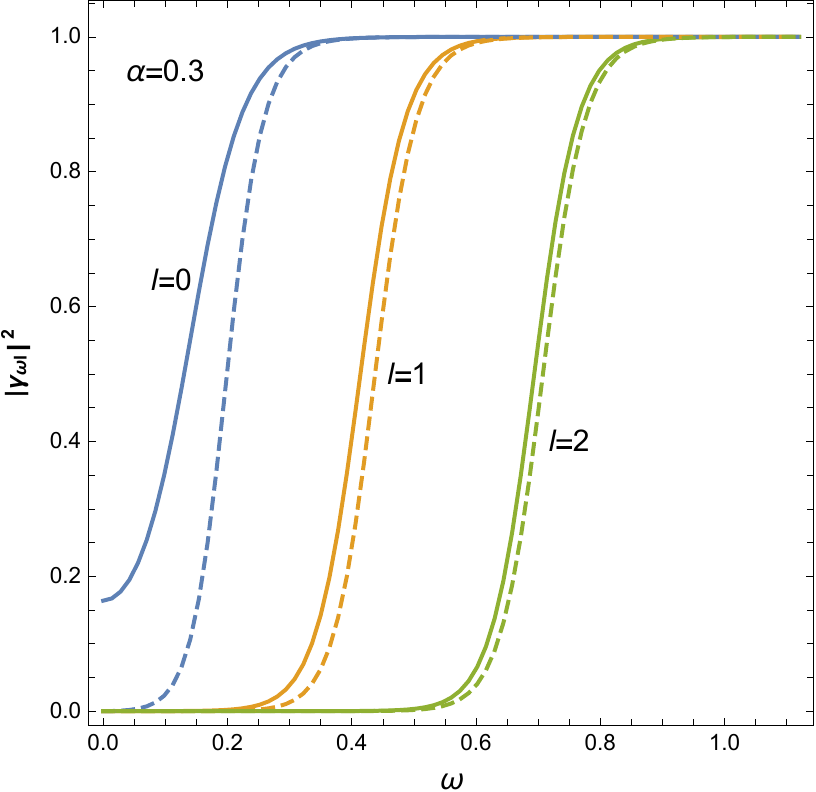}}{\footnotesize\par}
\par\end{centering}
{\footnotesize{}\caption{\label{fig:XiL} Effect of $\xi$ on the greybody factor for different
$l$. Solid lines for $\xi=0,$ dashed lines for $\xi=0.3$. We fix
$\Lambda=0.1,\alpha=0.3$ here.}
}{\footnotesize\par}
\end{figure}

Another interesting property of $\xi$ is that it introduces an effective
mass for the scalar, according to (\ref{eq:KG}). Thus for the dominant
mode $l=0$ with frequency $\omega=0$, it cannot cross over the effective
potential barrier and leads to a vanishing greybody factor. This can
be seen in Fig. \ref{fig:XiL}.

\subsection{\label{subsec:GreybodyLambda}Effect of $\Lambda$ on the greybody
factor }

\begin{figure}[h]
\begin{centering}
{\footnotesize{}}%
\begin{tabular}{ccc}
{\footnotesize{}\includegraphics[scale=0.6]{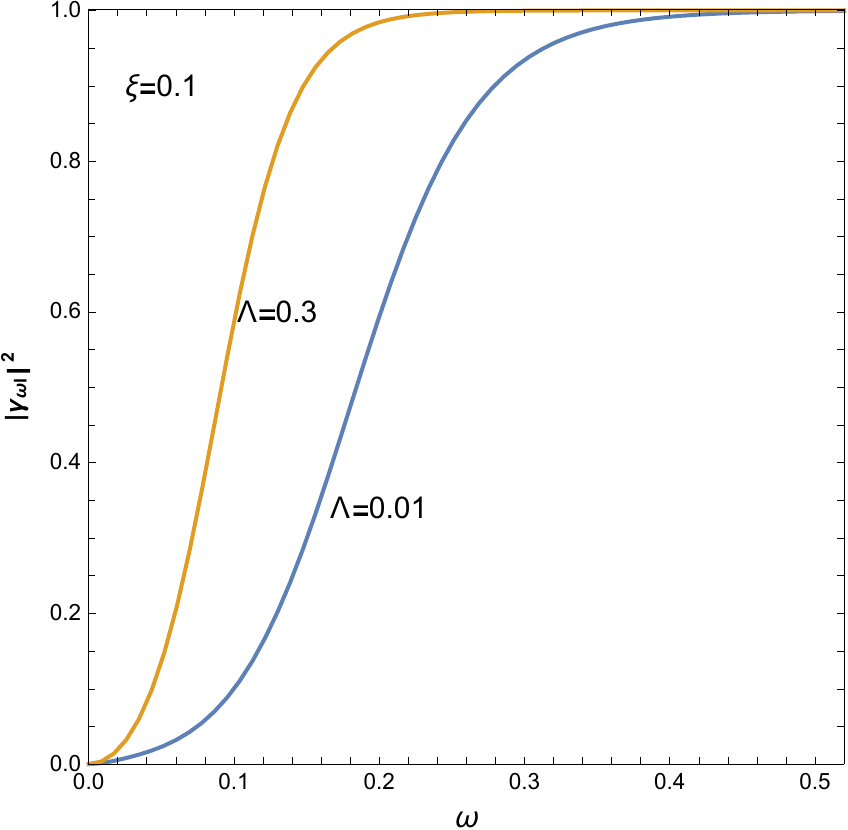}} & {\footnotesize{}\includegraphics[scale=0.6]{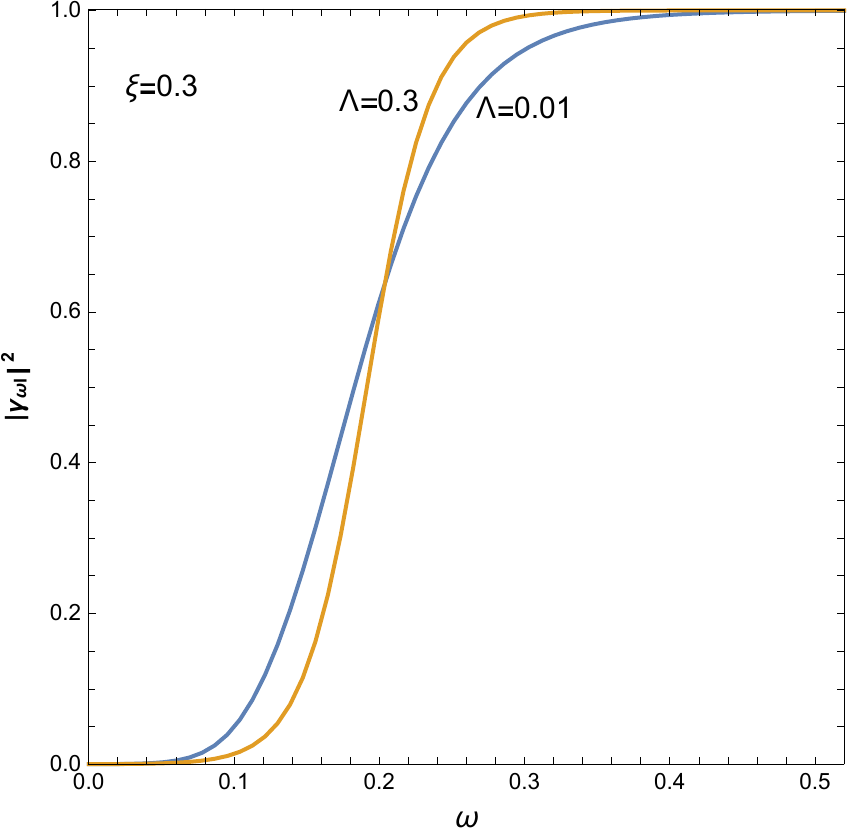}} & {\footnotesize{}\includegraphics[scale=0.6]{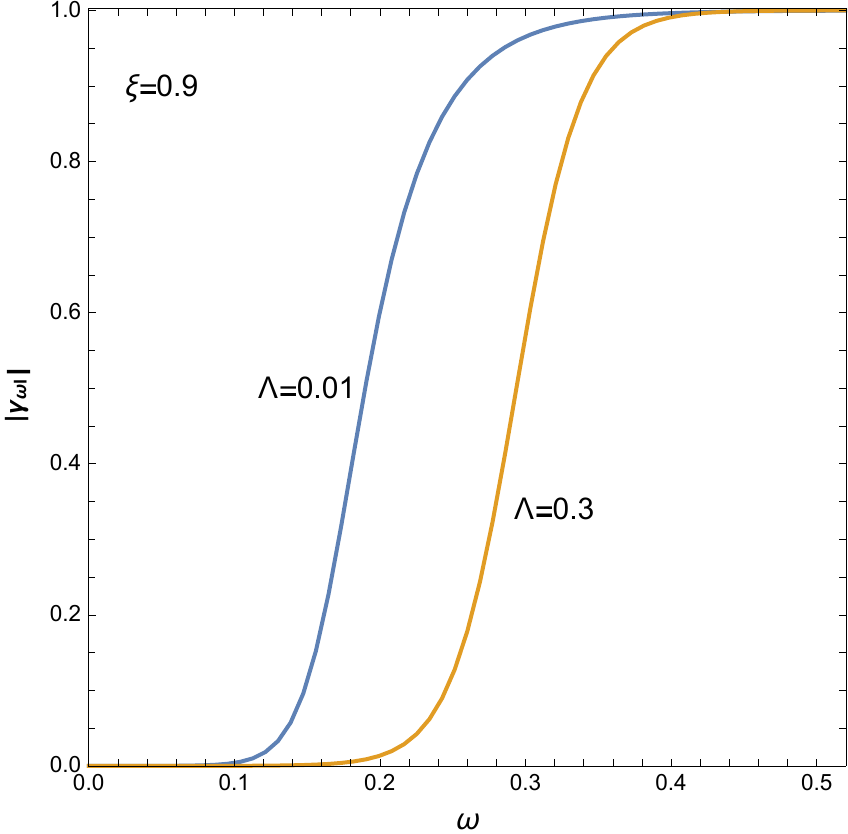}}\tabularnewline
\end{tabular}{\footnotesize\par}
\par\end{centering}
{\footnotesize{}\caption{\label{fig:AlphaXi} Effect of $\Lambda$ on the greybody factor.
We fix $l=0,\alpha=0.3$ here.}
}{\footnotesize\par}
\end{figure}

In the above subsection, we see that the cosmological constant plays
a subtle role for greybody factor. We study its effect on the greybody
factor in detail in this subsection. From Fig. \ref{fig:AlphaXi},
we see that for positive $\alpha$, when $\xi$ is small, $\Lambda$
increases the greybody factor. When $\xi$ is large, $\Lambda$ decreases
the greybody factor. Similar behavior has been observed in higher
dimensions \cite{Zhang2017,Li2019,Kanti2014,Pappas2016}. On the other
hand, $\Lambda$ can be understood as an homogeneously distributed
energy in the spacetime and boosts the particles to cross the effective
potential barrier. However, it also contributes to the effective mass
through (\ref{eq:KG}) and so hinders the particles to cross the barrier.
The first effect dominates when $\xi$ is small and the second effect
dominates when $\xi$ is large. When $\alpha$ is negative enough,
we find that $\Lambda$ increases the greybody factor in almost the
whole frequency region. The effects of $\Lambda$ on the effective
mass is always weaker.

{\footnotesize{}}

\section{The power spectra of Hawking radiation \label{sec:Energy-emission-rate}}

Once the greybody factor is worked out, the power spectra for Hawking
radiation can be obtained by definition \cite{Kanti2004,Harris2003,Kanti2005}
\begin{equation}
\frac{d^{2}E}{dtd\omega}=\frac{1}{2\pi}\sum_{l}\frac{N_{l}|\gamma_{\omega l}|^{2}\omega}{e^{\omega/T}-1},\label{eq:PowerSpectra}
\end{equation}
 where $T$ is the temperature of the black hole and $N_{l}=\frac{(2l+d-3)(l+d-4)!}{l!(d-3)!}$
is the multiplicity of the states that have the same $l$. As we have
learned in last section that the greybody factor of higher mode $l$
is non-vanishing only when the frequency is high enough. While the
power spectral is suppressed exponentially at high frequency according
to (\ref{eq:PowerSpectra}). Thus only the lower modes $l$ contribute
to the total power spectra significantly. We take $l\le6$ in the
numerical calculations hereafter.

{\footnotesize{}}
\begin{figure}[h]
\begin{centering}
{\footnotesize{}}%
\begin{tabular}{cc}
{\footnotesize{}\includegraphics[scale=0.8]{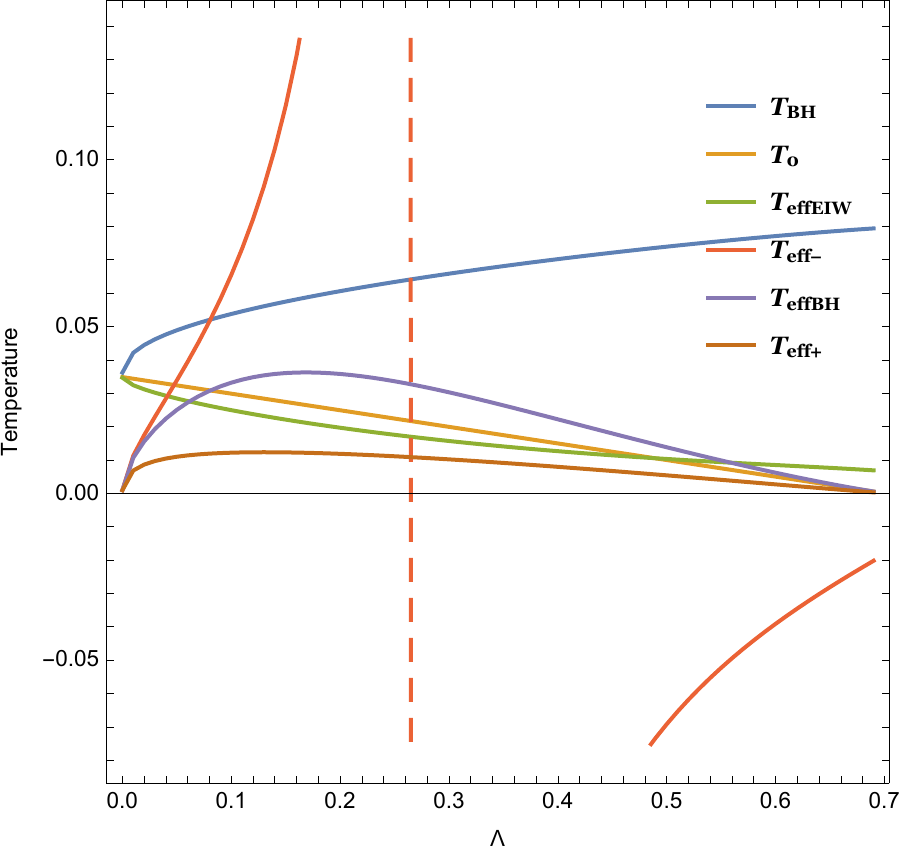}} & {\footnotesize{}\includegraphics[scale=0.8]{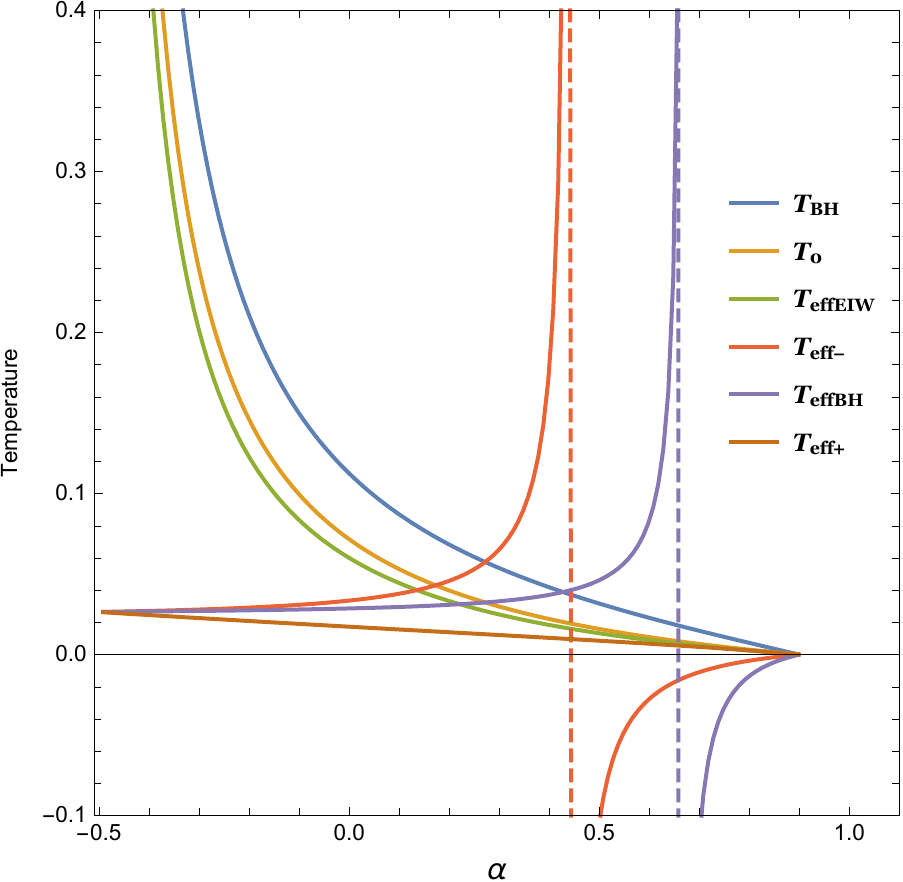}}\tabularnewline
\end{tabular}{\footnotesize\par}
\par\end{centering}
{\footnotesize{}\caption{\label{fig:temperature} The dependence of the effective temperature
of the black hole (\ref{eq:4metric}) on $\Lambda$ (left, with fixed
$\alpha=0.3$) and $\alpha$ (right, with fixed $\Lambda=0.1$). The
extremal black hole satisfies $\alpha+\Lambda=1$ and $\alpha>-0.5$.}
}{\footnotesize\par}
\end{figure}

For black holes in dS spacetime, the temperature of the system is
subtle. One can define the temperature $T_{0}=\frac{\kappa_{h}}{2\pi}$
on the event horizon and $T_{c}=-\frac{\kappa_{c}}{2\pi}$ on the
cosmological horizon. $T_{0}$ is different with $T_{c}$ in general
such that the system is not in equilibrium. Inspired from the black
hole thermodynamics, various effective temperatures were proposed
\cite{Pappas2016,Kanti2017,Pappas2017}, such as $T_{\text{eff-}}=\left(\frac{1}{T_{c}}-\frac{1}{T_{0}}\right)^{-1}$,
$T_{\text{eff+}}=\left(\frac{1}{T_{c}}+\frac{1}{T_{0}}\right)^{-1}$,
$T_{\text{effEIW}}=\frac{r_{h}^{4}T_{c}+r_{c}^{4}T_{0}}{(r_{h}+r_{c})(r_{c}^{3}-r_{h}^{3})}$
and $T_{\text{BH}}=\frac{T_{0}}{\sqrt{f(r_{0})}}$, $T_{\text{effBH}}=\left(\frac{1}{T_{c}}-\frac{1}{T_{BH}}\right)^{-1}$.
Here $r_{0}$ is position where $\partial_{r}f|_{r=r_0}=0$. Around
this position, the black hole attraction balances the cosmological
repulsion. Their dependence on $\Lambda$ is shown in the left panel
of Fig. \ref{fig:temperature}. We see that $T_{BH}$ increases with
$\Lambda$, while $T_{0},T_{\text{effEIW}}$ decreases with $\Lambda$
monotonically. $T_{\text{effBH}}$ and $T_{\text{eff+}}$ have a maximum.
$T_{\text{eff-}}$ diverges at $\Lambda=0.265$. The dependence of
the effective temperature on $\alpha$ is shown in the right panel
of Fig. \ref{fig:temperature}. Very interestingly, we find the three effective temperatures go to zero when $\alpha=-0.5$. The reason is that the temperature of the event horizon of the black hole $T_{BH}$ and the temperature defined on the event horizon $T_0$ approach infinity for $\alpha=-0.5$, that is to say, the black hole is unlimited hot for this case which is worthy of further study. Also, we see that $T_{BH},T_{0},T_{\text{effEIW}}$
and $T_{\text{eff+}}$ decrease with $\alpha$. It's worth taking a moment to flag the fact all the temperatures are continuous across $\alpha=0$ although the black hole for a positive and negative $\alpha$ seem different. In addition, we find $T_{\text{eff-}}$ diverges at $\alpha=0.443$ and $T_{\text{effBH}}$ diverges at $\alpha=0.668$.
To get reasonable result for the power spectra, we should abandon
$T_{\text{eff-}}$ and $T_{\text{effBH}}$. For both small $\Lambda$
and large $\Lambda$, $T_{\text{eff+}}$ tends to zero, which will
lead to vanishing power spectra according to (\ref{eq:PowerSpectra}).
This is unreasonable and thus we also abandon $T_{\text{eff+}}$.
It has been shown in higher dimension that only $T_{BH}$ leads to
significant radiation in the whole parameter space \cite{Li2019}.
In this section, we therefore take $T_{BH}$ to study the power spectra
of Hawking radiation.

\subsection{Effects of $\alpha$ and $\xi$ on the power spectra}

\begin{figure}[h]
\begin{centering}
{\footnotesize{}}%
\begin{tabular}{cc}
{\footnotesize{}\includegraphics[scale=0.8]{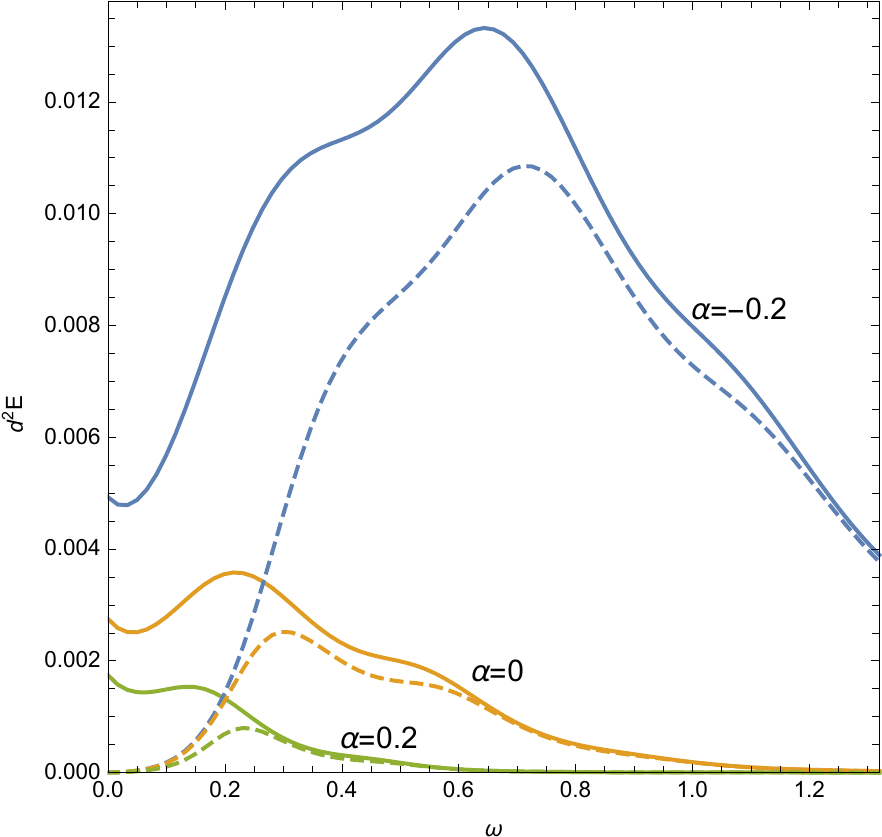}} & {\footnotesize{}\includegraphics[scale=0.8]{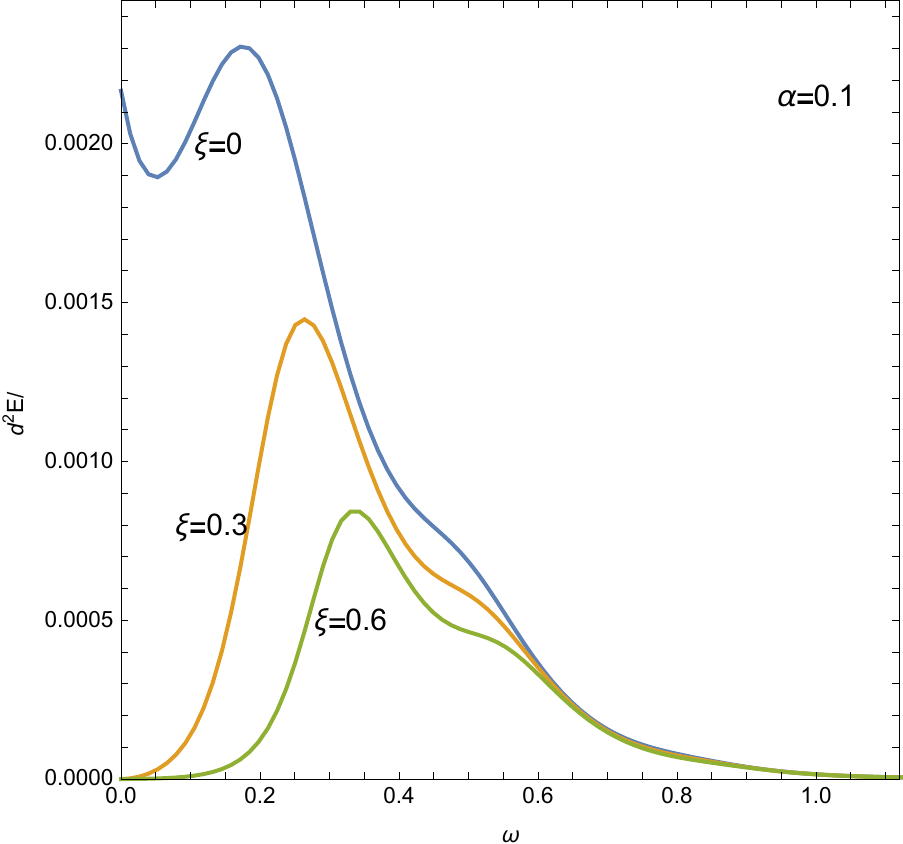}}\tabularnewline
\end{tabular}{\footnotesize\par}
\par\end{centering}
{\footnotesize{}\caption{\label{fig:PowerAlphaXi_a} The effects of $\alpha$ (left panel,
solid lines for $\xi=0$, dashed lines for $\xi=0.3$) and $\xi$
(right panel, with fixed $\alpha=0.1$) on the power spectra of Hawking
radiation. }
}{\footnotesize\par}
\end{figure}

The effects of $\alpha$ on the power spectra of Hawking radiation are shown in the left
panel of Fig. \ref{fig:PowerAlphaXi_a}. For both minimally and nonminimally
coupled scalar, the power spectra is suppressed as $\alpha$ increases,
which is contrast with its effect on the greybody factor. In particular, we find that line of the power spectra with $\alpha<0$ is above other lines  with $\alpha\ge0$. In other words, the intensity of Hawking radiation is much high with a negative $\alpha$, although the greybody factor is lower. The reason comes from that the temperature of the black hole is very high when $\alpha$ becomes negative. In fact, the temperature plays a more important role here. $T_{BH}$
decreases with $\alpha$, and the power spectra also decreases according
to (\ref{eq:PowerSpectra}). Note that for minimally coupled scalar,
the power spectra at low frequency is non-vanishing due to the finite
greybody factor there. For nonminimally coupled scalar, the power
spectra at low frequency tends to zero. The peak of the power spectra
moves to lower frequency as $\alpha$ increases, which is consistent
with Fig. \ref{subsec:XiL} where the greybody factor moves to the
left as $\alpha$ increases.

{\footnotesize{}}

The effects of $\xi$ on the power spectra is shown in the right panel
of Fig. \ref{fig:PowerAlphaXi_a}. The power spectra is suppressed
by $\xi$ in the whole frequency region. We have learned that in subsection
\ref{subsec:XiL}, the greybody factor can be enhanced at high frequency
when $\Lambda<3/56$. However, (\ref{eq:PowerSpectra}) tells us that
the power spectra is exponentially suppressed in high frequency. Thus
the qualitative behavior that $\xi$ suppresses the power spectra
is independent of $\alpha$. Note that the peak of the power spectra
for nonminimally coupled scalar moves to higher frequency when $\xi$
increases. This is consistent with Fig. \ref{fig:XiL} where the greybody
factor moves to the right as $\xi$ increases.

\subsection{Effect of $\Lambda$ on the power spectra  }

\begin{figure}[h]
\begin{centering}
{\footnotesize{}}%
\begin{tabular}{cc}
{\footnotesize{}\includegraphics[scale=0.8]{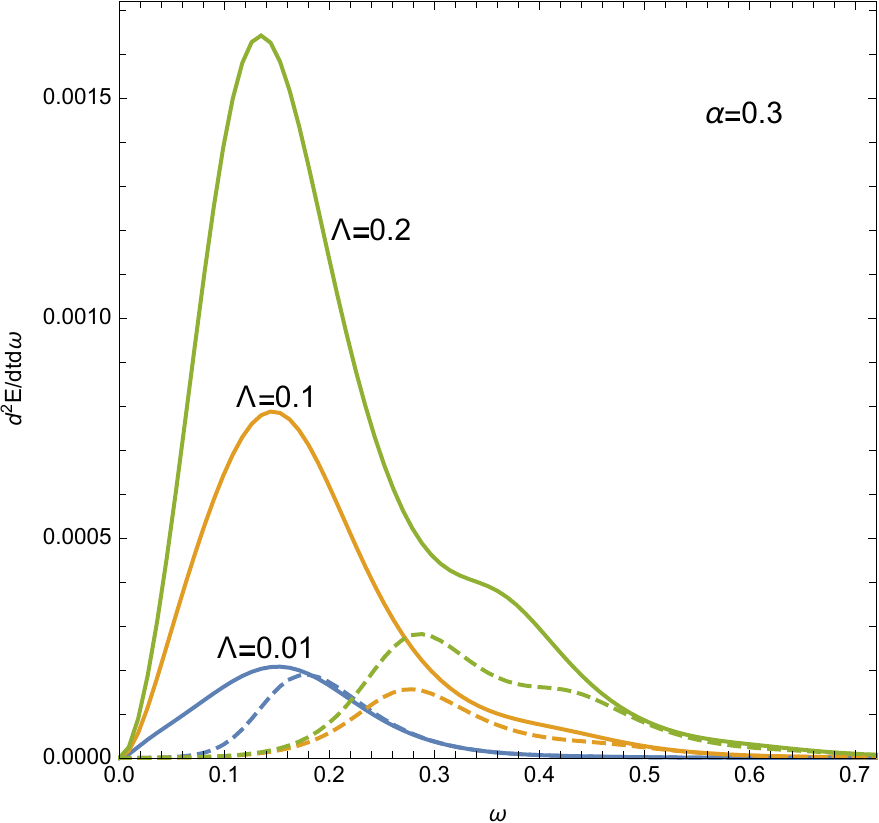}} & {\footnotesize{}\includegraphics[scale=0.8]{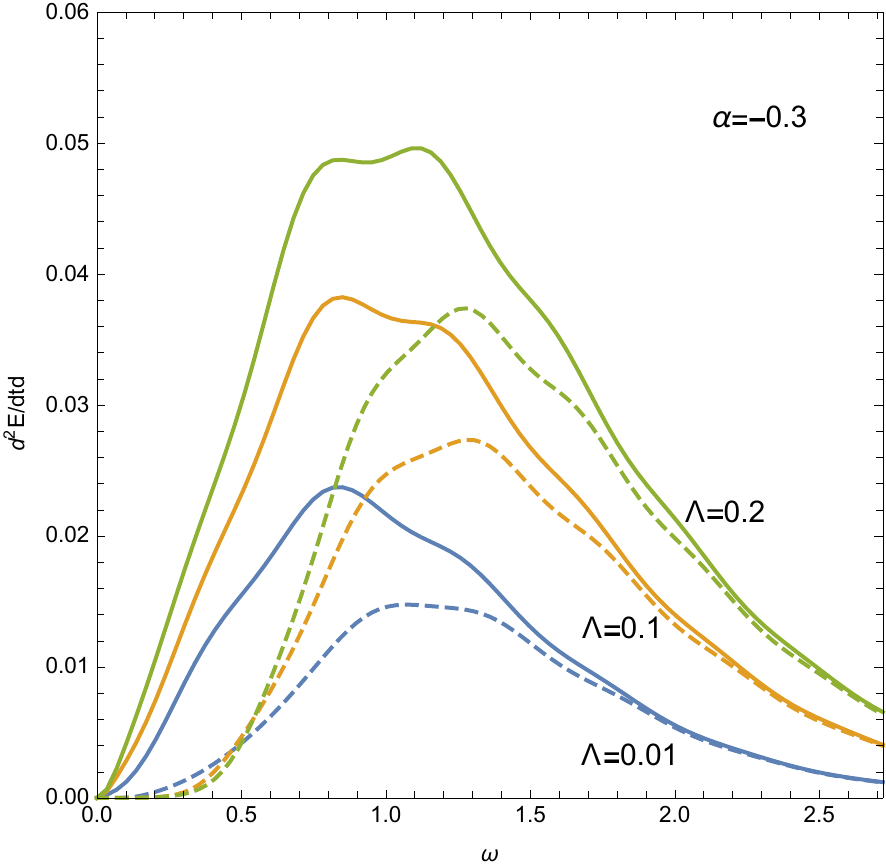}}\tabularnewline
\end{tabular}{\footnotesize\par}
\par\end{centering}
{\footnotesize{}\caption{\label{fig:PowerLambdaXi} Effect of $\Lambda$ on the power spectra
of Hawking radiation. Solid lines for $\xi=0.1$, dashed lines for
$\xi=0.6$. }
}{\footnotesize\par}
\end{figure}

The effect of $\Lambda$ on the power spectra of the Hawking radiation
is more subtle. For positive $\alpha$, as shown in the left panel
of Fig. \ref{fig:PowerLambdaXi}, the cosmological constant $\Lambda$
enhance the power spectra in the whole frequency region when $\xi$
is small. When $\xi$ is large, $\Lambda$ enhance the power spectra
only in the high frequency region. In the low frequency region, it
suppresses the power spectra., as shown by the dashed lines in Fig.
\ref{fig:PowerLambdaXi}. This behavior is consistent with Fig. \ref{fig:AlphaXi},
where the cosmological constant enhances the greybody factor when
$\xi$ is small and suppresses the greybody factor when $\xi$ is
large. However, when $\alpha$ is negative enough, the cosmological
constant enhances the power spectra in almost the whole frequency
region no matter how large $\xi$ is, as shown in the right panel
of Fig. \ref{fig:PowerLambdaXi}.

{\footnotesize{}}

\section{Summary \label{sec:Summary}}

The novel four dimensional Einstein-Gauss-Bonnet black holes found
recently have some distinct properties compared to their higher dimensional
companions. For example the spherically symmetric neutral 4D black
hole solutions in asymptotic dS spacetime can have three horizons,
while the solutions in higher dimensional spacetime have only two
horizons. On the other hand, we also found the spacetime contains a black hole when GB coupling constant is negative. Furthermore, the EGB black hole in 4D can be directly compared with the Schwarzschild black hole which is the most common model used to describe a real black hole in our universe. At this point, 4D EGB black hole has a significant advantage over the GB black holes in high dimensions. One thus expect that the Hawking radiation may also have
distinct properties compared to the higher dimensional case.

We studied the greybody factor of the Hawking radiation firstly. The
greybody factor is suppressed heavily by the angular momentum number
$l$ of the scalar mode. The Gauss-Bonnet coupling constant enhances
the greybody factor, while the nonminimally coupling constant $\xi$
of the scalar decreases it. In particular, we found compared to Schwarzschild-dS black hole ($\alpha=0$), the 4D EGB black hole with a negative $\alpha$ has a larger greybody factor.  When the frequency of the mode tends
to zero, the greybody factor vanishes for nonminimally coupled massless
scalar while has a finite value for minimally couple massless scalar.
The role of cosmological constant in greybody factor depends on $\xi$.
It enhances the greybody factor when $\xi$ is small, and decreases
it when $\xi$ is large. These behavior is similar to the higher dimensional
cases qualitatively, and can be understood intuitively from the viewpoint
of the effective potential.

We then studied the power spectra of the Hawking radiation of the
massless scalar in the 4D GBdS background. We analysed various definitions
of temperature in asymptotically dS spacetime and adopted the most
reasonable one to calculated the power spectra of the Hawking radiation.
We found that both $\xi$ and $\alpha$ suppress the power spectra. In need of special is that the power spectra with a negative $\alpha$ is over the cases with $\alpha\ge0$, which is to say the intensity of Hawking radiation with $\alpha<0$ is higher than others with $\alpha\ge0$, although the greybody factor is larger for $\alpha<0$.
The cosmological constant enhance the power spectra when $\xi$ is
small and suppress it when $\xi$ is large. The power spectra vanishes
for nonminimally coupled scalar while has a finite value for minimally
couples scalar.

The method used in this work can be developed to study the amplitude
of the superradiance of the charged black holes in 4D Einstein-Gauss-Bonnet
gravity. It is known that RN-dS black holes are linearly unstable
to spherical charged scalar perturbations \cite{Zhu2014}. It can
be expected that the charged 4D GB-(A)dS black holes have also superradiant
instability.

\section{Acknowledgments}

C.-Y. Zhang is supported by Natural Science Foundation of China under
Grant No. 11947067. MG and PCL are supported by NSFC Grant No. 11947210.
MG is also funded by China National Postdoctoral Innovation Program
2019M660278.

\end{document}